\documentstyle[twoside,fleqn,espcrc2,psfig]{article}

\newcommand{\AmS}{{\protect\the\textfont2
  A\kern-.1667em\lower.5ex\hbox{M}\kern-.125emS}}

\title{On the continuum limit of the Discrete Regge model in 4d}

\author{Elmar Bittner\address{Atominstitut der \"Osterreichischen Universit\"aten, A-1040 Vienna, Austria}$^,$\address{Institut f\"ur Theoretische Physik, Universit\"at Leipzig, D-04109 Leipzig, Germany}$^,$\thanks{Poster presented by E.~B. and  supported by Hoch\-schul\-jubil\"aums\-stiftung der Stadt Wien.},
Wolfhard Janke$^{\mbox{\scriptsize b}}$$^,$\thanks{W.~J. acknowledges partial
support by the EC IHP Network grant HPRN-CT-1999-00161: ``EUROGRID''.} and
Harald Markum$^{\mbox{\scriptsize a}}$
}
\begin{document}
\begin{abstract}
The Regge Calculus approximates a continuous manifold by a simplicial lattice,
keeping the connectivities of the underlying lattice fixed and taking the edge
lengths as degrees of freedom. The Discrete Regge model employed in this work
limits the choice of the link lengths to a finite number.
This makes the computational evaluation of the path integral much faster.
A main concern in lattice field theories is the existence of a continuum
limit which requires the existence of a continuous phase transition. The
recently conjectured second-order transition of the four-dimensional Regge 
skeleton at negative gravity
coupling could be such a candidate. We examine this regime with Monte Carlo
simulations  and
critically discuss its behavior.
\end{abstract}

\maketitle

\section{REGGE QUANTUM GRAVITY}
One promising method to quantize the theory of gravitation employs 
the Euclidean path integral
\begin{equation} \label{Z}
Z=\int D[q] e^{-I(q)} ~,
\end{equation} 
where the partition function 
describes
a fluctuating space-time manifold. In the Regge approach the
(quadratic) link lengths $q_l$ represent the dynamical degrees of freedom
\cite{hart}, deforming continuously a simplicial lattice with fixed incidence 
matrix, whereas in the somehow complementary approach of dynamical triangulation
the incidence matrix is fluctuating with constant link lengths.
Any of these two approaches is plagued with various problems; for 
the Regge approach see \cite{herb,we}.

In ``conventional'' Regge theory the Regge-Einstein action including a cosmological 
term \cite{regge},
\begin{equation} \label{I_R}
I_r=-\beta\sum\limits_t A_t\delta_t + \lambda\sum\limits_s V_s ~,
\end{equation}
is used.
The first sum runs over all products of triangle area $A_t$ times
corresponding deficit angle $\delta_t$ weighted by the bare gravitational 
coupling $\beta$. The second sum extends over the volumes $V_s$ of the
4-simplices of the lattice and allows together with the cosmological
constant $\lambda$ to set an overall scale in the action.

The Discrete Regge model was invented in an attempt to reformulate
(\ref{Z}) as the partition function of a spin system \cite{2d,mark}. 
It is defined by restricting the squared link lengths to take on only 
two values
\begin{equation} \label{q_l}
q_l=b_l(1+\epsilon\sigma_l) ~,\quad \sigma_l\in Z_2 ,
\end{equation}
where  $b_l=1,2,3$, and $4$ for edges, face diagonals, body diagonals and the 
hyperbody diagonal of a hypercube, respectively, is chosen to allow for fluctuations around flat space. 
The Euclidean triangle inequalities are fulfilled automatically  as long as 
$\epsilon$ is smaller than a maximum value $\epsilon_{\rm max}$.
The measure  $D[q]$ in the quantum gravity path integral
is taken to unity for all possible link configurations. 

Numerical simulations of the $Z_2$ system become extremely efficient by
implementing look-up tables and a heat-bath algorithm.
In this work typically $200\,000 - 500\,000$ iterations have been generated. 
Calculations have been performed with the parameter $\epsilon=0.0875$ 
and the cosmological constant $\lambda=0$ because (\ref{q_l}) 
already fixes the average lattice volume.

\section{RESULTS}
The Discrete Regge model -- like full Regge theory -- exhibits two phase transitions \cite{mel}. One is located
at a negative value and  the other one at a positive value of the bare gravitational coupling.
Although earlier work concentrated on the latter transition, there is no reason for favoring
a positive value of $\beta$ over a negative one. Since a Wick rotation from the Lorentzian
to the Euclidean sector of quantum gravity
is not feasible in general, the sign in the exponential of the path integral
is not fixed a priori.

\begin{figure}[t]
\psfig{figure=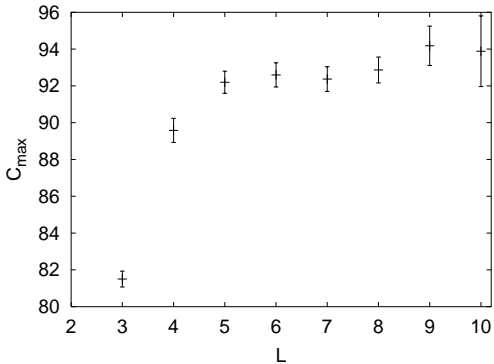,width=7cm,height=5cm}
\vspace{-8mm}
\caption{\label{fig1} FSS of the specific-heat maxima $C_{\rm max}$ as a 
function of the lattice size $L$.}
\vspace{-4mm}
\end{figure}

The transition for positive gravitational coupling has been observed  before the one
at negative $\beta$ \cite{hamber}.
There, histograms, e.g. of $A_t\delta_t$ \cite{mel}, clearly show a two-peak
structure.  The two phases coexist and tunneling from one phase to the other
and back takes place. The system also exhibits a hysteresis; the transition occurs
at a larger value of $\beta$ if the
simulation is started from a configuration in the well-defined phase than it does if the calculation
is started from a ``frozen'' configuration. Given all these pieces of evidence, we 
conjecture that the transition at $\beta > 0$
is of first order.

To determine the order of the transition at negative $\beta$ we employed
in Ref.~\cite{mel} histogram techniques 
and  used the Binder--Challa--Landau (BCL) cumulant criterion \cite{B1}.
The BCL cumulant is defined as
$B_L := 1- { \langle E^4 \rangle \over 3 \langle E^2 \rangle ^2 }$, 
with $E$ being the action of the system under consideration.
It was evaluated for  $A_t \delta_t$ on different lattice sizes with $L=3$
to $10$ vertices per direction, simulating the system at several values of
the bare coupling $\beta$ with high statistics.
For the BCL cumulant a trend towards 2/3 was observed and all histograms 
showed a clear one-peak structure, cf.~\cite{mel}.

In the present Monte Carlo simulations 
we recorded for every run the time series of the energy density
$e=E/N_0$ and the magnetization density $m= \sum_l \sigma_l /N_0$, with the lattice size
$N_0=L^4$.
To obtain results for the various observables ${\cal O}$ at values of the bare gravitational
coupling $\beta$ in an interval around the simulation point $\beta_0$, we applied the
reweighting method~\cite{Ferr}.

\begin{figure}[t]
\psfig{figure=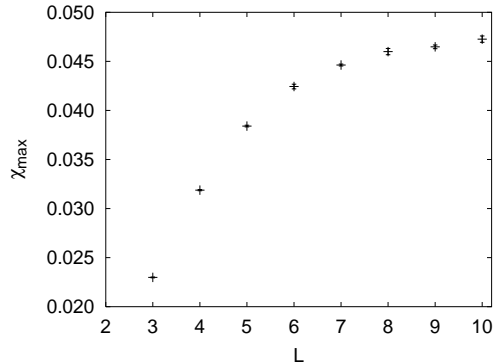,width=7cm,height=5cm}
\vspace{-8mm}
\caption{\label{fig2}  FSS of the susceptibility maxima $\chi_{\rm max}$,
indicative for a cross-over.}
\vspace{-4mm}
\end{figure}

With the help of the time series we can compute the specific heat,
\begin{equation}
C(\beta)=\beta^2 N_0 (\langle e^2 \rangle- \langle e\rangle^2)~,
\end{equation}
and the (finite lattice) susceptibility,
\begin{equation}
\chi(\beta)=N_0(\langle m^2 \rangle -\langle |m| \rangle^2)~.
\end{equation}
Figures \ref{fig1} and \ref{fig2} show the finite-size scaling (FSS) of
the maxima of the specific heat $C_{\rm max}$ and the susceptibility
$\chi_{\rm max}$, respectively. While the behavior of
$C_{\rm max}$ could still be explained as critical scaling with a negative 
exponent $\alpha$, the flattening of $\chi_{\rm max}$ is rather unusual for
a second-order transition and should be taken as an indication for a 
cross-over regime. 
 
\begin{figure}[t]
\psfig{figure=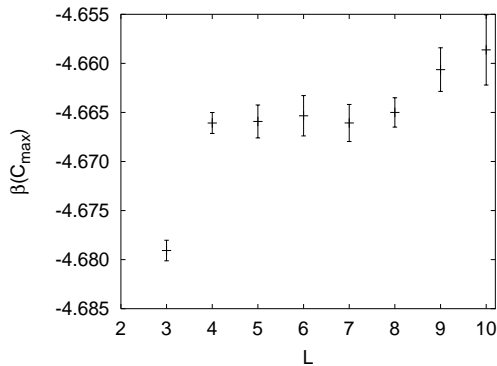,width=7cm,height=5cm}
\vspace{-8mm}
\caption{\label{fig3} Volume dependence of the critical gravitational coupling
$\beta(C_{\rm max})$, as determined from the specific-heat maxima $C_{\rm max}$
in Fig. \ref{fig1}.}
\vspace{-4mm}
\end{figure}

\begin{figure}[t]
\psfig{figure=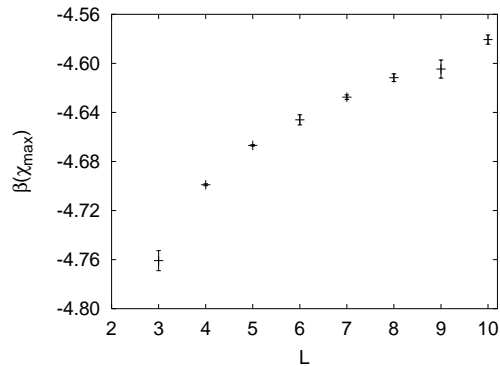,width=7cm,height=5cm}
\vspace{-8mm}
\caption{\label{fig4} Volume dependence of the critical gravitational coupling
$\beta(\chi_{\rm max})$, as determined from the susceptibility maxima
$\chi_{\rm max}$
in Fig. \ref{fig2}.}
\vspace{-4mm}
\end{figure}

Another feature of the system can be seen in Figs. \ref{fig3} and \ref{fig4}
depicting the critical gravitational coupling, as determined from the
specific-heat maxima $C_{\rm max}$ and the susceptibility maxima 
$\chi_{\rm max}$.
In the case of a second-order transition the extrapolations of all 
pseudo-transition points lead to one infinite-volume critical value. 
This seems to be violated in the four-dimensional Discrete Regge model and 
thus favors the interpretation as a cross-over phenomenon over a true,
thermodynamically defined phase transition.

\section{CONCLUSION}

In the present analysis of the scaling of the maxima of the specific heat 
and susceptibility we found evidence for a cross-over regime in the 
four-dimensional Discrete 
Regge model of quantum gravity in the negative coupling region. If this can be
substantiated by further investigations and also for the Regge theory with
continuous link lengths, the existence of a continuum limit at negative
bare gravitational coupling is questionable. 
This is of major concern
for a continuum theory of quantum gravity with matter fields~\cite{prd02}.

\begin{thebibliography}{9}
\bibitem{hart} J. Hartle, J. Math. Phys. 26 (1985) 804; 27 (1986) 287;
 30 (1989) 452.
\bibitem{herb} H.W. Hamber, Phys. Rev. D45 (1992) 507;
Nucl. Phys. B400 (1993) 347.
\bibitem{we} W. Beirl, E. Gerstenmayer, H. Markum and J. Riedler, Phys.
 Rev. D49 (1994) 5231.
\bibitem{regge} T. Regge, Nuovo Cimento 19 (1961) 558.
\bibitem{2d}
 W. Beirl, H. Markum and J. Riedler, Int. J. Mod. Phys. C5 (1994) 359;
 W. Beirl, P. Homolka, B. Krishnan, H. Markum and J. Riedler, Nucl.
 Phys. B (Proc. Suppl.) 42 (1995) 710.
\bibitem{mark}
 T. Fleming, M. Gross and R. Renken, Phys. Rev. D50 (1994) 7363.
\bibitem{mel} W. Beirl, A. Hauke, P. Homolka, B. Krishnan, H. Markum
and J. Riedler, Nucl. Phys. B (Proc. Suppl.) 47 (1996) 625;
J. Riedler, W. Beirl, E. Bittner, A. Hauke, P. Homolka and H. Markum,
Class. Quant. Grav. 16 (1999) 1163.
\bibitem{hamber}
H.W. Hamber in: {\em Proceedings of the 1984 Les Houches Summer School,
Session XLIII}, edited by K. Osterwalder and R. Stora
(North Holland, Amsterdam, 1986);
H.W. Hamber and R.M. Williams, Phys. Lett. B157 (1985) 368;
H.W. Hamber, Phys. Rev. D61 (2000) 124008.
\bibitem{B1} K. Binder and D.P. Landau, Phys. Rev. B30 (1984) 1477;
M.S.S. Challa, D.P. Landau and K. Binder, Phys. Rev. B34 (1986) 1841.
\bibitem{Ferr} A.M. Ferrenberg and R.H. Swendsen, Phys. Rev. Lett. 61
(1988) 2635; 63 (1989) 1195; 1658.
\bibitem{prd02}
E. Bittner, W. Janke and H. Markum,
Phys. Rev. D66 (2002) 024008; Acta Physica Slovaca 52 (2002) 241.
\end{thebibliography}
\end{document}